\documentclass[twocolumn,floatfix,showpacs,prd,aps,tightenlines]{revtex4}
\usepackage{graphicx}
\usepackage{psfrag}
\usepackage{dcolumn}
\usepackage{bm}

\newcommand{\bea}{\begin{eqnarray}}
\newcommand{\eea}{\end{eqnarray}}
\newcommand{\beq}{\begin{equation}}
\newcommand{\eeq}{\end{equation}}

\begin{document}

\title{The last orbit of binary black holes}

\author{M. Campanelli}  \affiliation{Department of Physics and Astronomy,
and Center for Gravitational Wave Astronomy,
The University of Texas at Brownsville, Brownsville, Texas 78520}

\author{C. O. Lousto} \affiliation{Department of Physics and Astronomy,
and Center for Gravitational Wave Astronomy,
The University of Texas at Brownsville, Brownsville, Texas 78520}

\author{Y. Zlochower} \affiliation{Department of Physics and Astronomy,
and Center for Gravitational Wave Astronomy,
The University of Texas at Brownsville, Brownsville, Texas 78520}

\date{\today}

\begin{abstract}
We have used our new technique for fully numerical evolutions of
orbiting black-hole binaries without excision to model the last orbit
and merger of an equal-mass black-hole system. We track the
trajectories of the individual apparent horizons and find that the
binary completed approximately one and a third orbits before forming a
common horizon. Upon calculating the complete gravitational radiation
waveform, horizon mass, and spin, we find that the binary radiated
$3.2\%$ of its mass and $24\%$ of its angular momentum. The early part
of the waveform, after a relatively short initial burst of spurious
radiation, is oscillatory with increasing amplitude and frequency, as
expected from orbital motion. The waveform then transitions to a
typical `plunge' waveform; i.e.\ a rapid rise in amplitude followed by
quasinormal ringing. The plunge part of the waveform is remarkably
similar to the waveform from the previously studied `ISCO'
configuration. We anticipate that the plunge waveform, when starting
from quasicircular orbits, has a generic shape that is essentially
independent of the initial separation of the binary.

\end{abstract}

 \maketitle

\section{Introduction}\label{Sec:Intro}

The study of the late orbital stage of black-hole binaries is of
particular interest because they are thought to be the most likely
sources detected by gravitational wave observatories such as
LISA~\cite{Danzmann:2003tv} and LIGO~\cite{LIGO} (which is now
reaching its design sensitivity).  Besides, the two body problem in
General Relativity is, in itself, a genuinely interesting theoretical
problem.  Studies of families of binary-black-hole initial data in
quasicircular orbits set the periods of the innermost stable circular
orbit (ISCO) to $37M$ for the Bowen-York family of initial data
\cite{Cook91,Baumgarte00a}, $57M$ for the Thin-Sandwich
family~\cite{Cook:2004kt,Hannam:2005rp}, and $49M$ from third-order
post-Newtonian approximation (3PN)~\cite{Blanchet:2001id}.  Evolution
of binary black holes from these locations lead to plunge motion
\cite{Baker:2001nu,Baker:2002qf,Baker:2004wv}; performing a fraction of
an orbit before a common horizon encompasses the two black holes 
(i.e. forming a single, distorted black holes).

There have been several remarkable advancements in Numerical
Relativity in the past few years, and, particularly since the work of
\cite{Pretorius:2005gq,Campanelli:2005dd,Baker:2005vv}, it now seems
possible to evolve orbiting black-hole binaries out from arbitrary
distances to the merger and ringdown. The first major breakthrough in
numerical evolutions of these systems was reported by Br\"ugmann et.\
al.~\cite{Bruegmann:2003aw}.  Using a careful choice of corotating shift
and singularity excision, they were able to evolve a black-hole
binary, starting from initial data for a quasicircular binary, for
more than a complete orbit. Their work was recently
verified by Diener et.\ al.~\cite{Diener:2005mg}. However, in both
cases it was not possible to extract the waveform from the merger.
The first fully numerical evaluation of the waveform from an orbiting
black-hole binary was reported by Pretorius~\cite{Pretorius:2005gq}.
Pretorius evolved a system in which two scalar fields collapsed to
form individual black holes, which then formed a merging elliptical
binary. The evolution used a direct discretization of a second
order in time system with singularity excision. Recently, a new
technique~\cite{Campanelli:2005dd, Baker:2005vv} has been developed
for evolving black-hole binaries that uses the more conventional
BSSN~\cite{Nakamura87,Shibata95, Baumgarte99} system of equations
(which are first-order in time). This new technique is based on the
puncture approach, but allows the singular punctures to move across
the grid.  Singularity excision is not required and the
new system has the advantages that it is very easy to implement and
appears to be very accurate. This is explicitly demonstrated 
in Ref.~ \cite{Herrmann:2006ks} where this technique was applied 
to study the merger kicks of unequal-mass black hole binaries 
verifying those previously computed in Ref. \cite{Campanelli:2004zw}.

We use this new `moving puncture' approach to evolve the same initial
configuration as in~\cite{Bruegmann:2003aw} and confirm that the
system does indeed undergo more than a full orbit before a common
horizon forms.  We were also able to accurately extract the waveform
and final horizon parameters.  We find very good agreement in the
radiated energy, angular momentum, and merger time between those
calculated from the horizon properties and from the waveform.

\section{Formulation}\label{Sec:Form}
Our simulations of orbiting black-hole binaries are based on a
modification to the standard puncture approach.  In the puncture
approach~\cite{Brugmann:1997uc} the metric on the initial slice is
given by~\cite{Brandt97b}
$
\gamma_{ab} = (\psi_{BL} + u)^4 \delta_{ab},
$
where
$
\psi_{BL} = 1 + \sum_{i=1}^n m_i/(2 r_i)
$
is the Brill-Lindquist conformal factor, $m_i$ is the mass parameter
of puncture $i$, $r_i$ is the coordinate distance to puncture $i$, and
$u$ is finite on the punctures.

In the standard puncture approach the locations of the punctures are
fixed (one imposes that the shift vanishes at the puncture location),
and the singular behavior (i.e.\ $\psi_{BL}$) in the metric is handled
analytically. A consequence of fixing the punctures is that the
coordinates become highly distorted, and this, in turn, causes
numerical runs of orbiting black holes (without excision or
corotation) to crash relatively early.  In a recent paper we
introduced a new technique for evolutions with moving
punctures~\cite{Campanelli:2005dd} (see~\cite{Baker:2005vv} for an
alternative implementation). This new technique, which is based on the
BSSN formulation of General Relativity \cite{Nakamura87,Shibata95,
Baumgarte99}, does not require either excision or a corotating shift.
Our technique replaces the BSSN conformal exponent $\phi$, which is
infinite on the punctures, with the $C^4$ field $\chi = \exp(-4\phi)$.
This new variable, along with the other BSSN variables, will remain
finite provided that one uses a suitable choice for the gauge.

We obtained accurate, convergent waveforms by evolving this system in
conjunction with a modified 1+log lapse, a modified Gamma-driver shift
condition~\cite{Alcubierre02a,Campanelli:2005dd}, and an initial lapse
set to $\alpha=\psi_{BL}^{-2}$.  The lapse and shift are evolved with
\begin{eqnarray}
  \partial_0 \alpha &=& - 2 \alpha K, \\
  \partial_t \beta^a &=& B^a,
   \quad \partial_t B^a = 3/4 \partial_t \tilde \Gamma^a - \eta B^a.
\end{eqnarray}

We use the {\it LazEv} framework~\cite{Zlochower:2005bj} to
numerically evolve this new system. Unlike in the fixed puncture
approach, we do not reduce the order of finite differencing near the
punctures. We use the standard centered fourth-order stencils for all
derivatives except for the advection terms (i.e.\ terms of the form
$\beta^i \partial_i$) where we use upwinded fourth-order
stencils. These stencils were modified near the boundary. At the
second point from the boundary we use fourth-order centered stencils
for all derivatives, and at the first point from the boundary, we use
second-order centered stencils.  We use the standard fourth-order
Runge-Kutta algorithm for the time evolution and radiative boundary
conditions for all evolved variables.

\section{Initial Configuration}\label{Sec:Config}
Following~\cite{Bruegmann:2003aw} we choose black hole initial data
from a quasicircular sequence~\cite{Tichy:2003qi} with parameters
\begin{eqnarray*}
&&m/M=0.47656,\quad P/M=0.13808,\quad Y/M=\pm3.0,\nonumber \\
&&L/M=9.32,\quad J/M^2=0.82843,\quad M\Omega=0.054988,
\end{eqnarray*}
where $m$ is the mass of each single black hole, $M=1$ provides the scale,
$P$ is the magnitude of the linear
momenta (equal but opposite and perpendicular to the line connecting
the holes), $(0,Y,0)$ is the coordinate
location of the punctures, $L$ is the proper distance
between the apparent horizons, $J$ is the total angular momentum,
and $\Omega$ is the orbital frequency.
We use the Brandt-Br\"ugmann approach along with the
BAM\_Elliptic~\cite{Brandt97b,cactus_web} Cactus thorn to solve
for these initial data. The total ADM mass for this configuration is
$M_{ADM} = 0.98461 M$.

The initial choice for the lapse is $\alpha = \psi_{BL}^{-2}$ and the
initial choice for the shift is $\beta^i = B^i = 0$.

We evolved these data with grid resolutions of $M/21$, $M/24$, and
$M/27$; and gridsizes of $224^2\times 448$, $256^2\times512$, and
$288^2\times576$ respectively (we exploited the $\pi$-rotation
symmetry about the $z$ axis and reflection symmetry about the
equatorial plane to halve the number of gridpoints in the $x$ and $z$
directions).  We used a `multiple transition' Fisheye
transformation, which is an extension of the `transition' Fisheye 
transformation~\cite{Baker:2001sf,Alcubierre02a}, to place the boundaries at
$114M$. 
The `multiple transition' transformation has the form
$R = C\, r$, where $R$ is the physical radius 
corresponding to the coordinate radius $r$,
\begin{eqnarray}
C &=& a_n +\sum_{i=1}^{n} \kappa_i/r 
  \log\left(\frac{\cosh((r+r0_i)/s_i)}
  {\cosh((r-r0_i)/s_i)}\right), \\
\kappa_{i} &=& \frac{(a_{i-1} - a_i)  s_i }{2\, \tanh(r0_i/s_i)},
\end{eqnarray}
$n$ is the number of transitions, $a_i$ is the deresolution parameter
in region $i$, $a_0$ is the central resolution, $r0_i$ is the center 
of the $i$th transition, and $s_i$ is the width of the $i$th transition.
For these runs we used the parameters,
$n=2$, $a_0 = 1$, $a_1 = 5$, $a_2=30$, $r0_1 = 5$, $r0_2 = 7.5$,
$s_1 = s_2 = 0.75$.
We also evolved these data with a resolution of $M/21$ and a
gridsize of $288^2\times576$, which placed the boundary at $176.6M$,
to quantify the dependence of the results on the location of the
boundary. For this run we used the Fisheye parameters,
$n=1$, $a_0=1$, $a_1=25$, $r0_1 = 7.0$, $s_1 = 25$.

\section{Results}\label{Sec:results}

We used Jonathan Thornburg's AHFinderDirect
thorn~\cite{Thornburg2003:AH-finding} to find apparent horizons.  The
(coordinate) time of first appearance of the common horizon depends on
resolution, with $T_{cah} = 111M$ for the $M/21$ run, $T_{cah} = 113M$
(estimated) for $M/24$, and $T_{cah}=114.3$ (estimated) for the $M/27$
run.  An extrapolation of these data to infinite resolution puts the
appearance of the first common horizon at $T_{cah} = 125M$. The common
horizon has an irreducible mass of $M_{irr}=(0.8848\pm0.0002) M$ and
specific spin (measured from the distortions of the horizon) of
$\tilde a = J_{\cal H}/M_{\cal H}^2= 0.688\pm0.001$.  These parameters
correspond to a horizon mass of $M_{\cal H} = (0.952\pm0.002) M$ and
angular momentum $J_{\cal H}=(0.6232\pm0.003) M^2$. Hence
$(3.3\pm0.2)\%$ of the mass and $(24.7\pm0.4)\%$ of the angular
momentum are radiated away.

\begin{figure}
\begin{center}
\includegraphics[width=3.3in]{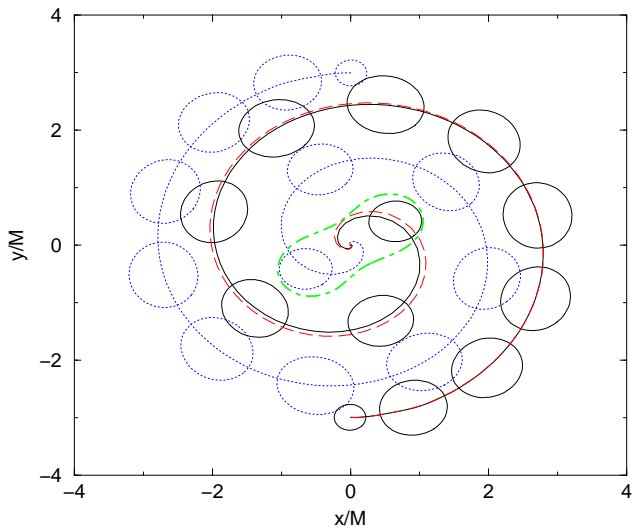}
\caption{The puncture trajectories and apparent horizon profiles on
the $xy$ plane for the $M/21$ run. The solid and dotted spirals are
the puncture trajectories, the solid and dotted ellipsoids are the
individual apparent horizons (at every 10M of evolution), the dot-dash
`peanut shaped' figure is the first detected common horizon, and the
dashed spiral is the puncture trajectory for the $M/27$ run. The
initial growth of the individual apparent horizon is due to the
non-ideal (vanishing) initial data for the shift.  Note that we track
the puncture positions throughout the evolution. The period of the
last orbit is around $62M$.  The last orbit begins when the punctures
are located at $2.6M$ from the origin.  }
\label{fig:track_wth_ah}
\end{center}
\end{figure}

Figure~\ref{fig:track_wth_ah} shows the tracks of the punctures, the
individual horizons every $10M$ of evolution, and the first common
horizon. The plot was generated using the $M/21$ resolution run with
the boundaries at $176.6M$. In addition the plot shows the puncture
trajectory for the $M/27$ resolution run. Note that the binary
completes one and a third orbits before the common horizon forms
(although we caution the reader that tracks are gauge dependent). The
period of the last orbit is approximately $62M$. The puncture
trajectories were calculated by integrating $\partial_t x^i_{punct} =
- \beta^i_{punct}$, where $\beta^i_{punct}$ is the interpolated value
of the shift on the puncture (the puncture never lies on a
gridpoint). The last orbit begins when the punctures are located at
$2.6M$ from the origin in the coordinate conformal space.
 
\begin{figure}
\begin{center}
\includegraphics[width=3.3in]{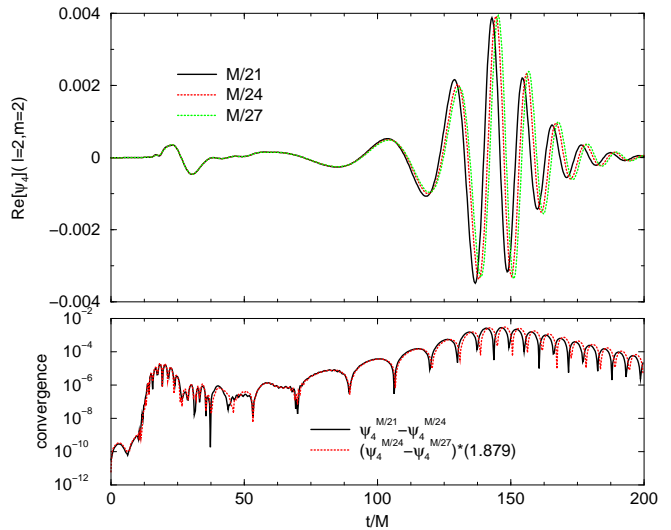}
\caption{The $(\ell=2,m=2)$ mode of $\psi_4$ at $r=20M$. The top
plot shows the waveforms  for central resolutions of 
$M/21$, $M/24$, and $M/27$. The bottom plot shows the differences 
$\psi_4(M/21) - \psi_4(M/24)$ and $\psi_4(M/24) - \psi_4(M/27)$, with
the latter rescaled by $1.879$ to demonstrate fourth-order convergence.
Note the spurious radiation around $t=26M$.}
\label{fig:psi4}
\end{center}
\end{figure}

We use the Zorro thorn~\cite{Baker:2001sf,Zlochower:2005bj} to
calculate $\psi_4$ and decompose it into spherical harmonics of spin
weight $-2$.
The two dominant modes are $(\ell=2,m=+2)$ and $(\ell=2, m=-2)$, where
the coefficient of the two modes are complex
conjugates. Figure~\ref{fig:psi4} shows the $(\ell=2,m=2)$ mode of
$\psi_4$ at $r=20M$ for the three resolutions and a convergence plot
of these data.  The waveforms converge to fourth-order up to $t=141M$
(the convergence rate past $t=141M$ is obscured by the large phase
error, but is better than second-order).  The oscillation in $\psi_4$
around $t=26M$ is due to spurious radiation in the initial data. This
spurious radiation quickly leaves the system and becomes smaller than
the orbital waveform at $t\sim50M$. Hence the radiation from the last
orbit, which begins at $t\sim50M$ as seen by our observer located at
the radial coordinate $r=20M$ (see Figure~\ref{fig:track_wth_ah}), is
essentially uncontaminated.  We find that the $(\ell=2,m=2)$
quasinormal mode, for the $M/21$ resolution run with distant
boundaries, has a frequency of $M_{\cal H}\,\omega/\alpha_{20} =
0.549\pm0.001$ ($\alpha_{20}= 0.954$ is the average value of the lapse
at $r=20M$ at late times). The reported error is from the fit to a
damped sinusoidal function and does not include finite difference
errors. This frequency corresponds~\cite{Echeverria89} to a specific
spin of $\tilde a = J_{\cal H}/M_{\cal H}^2 = 0.673\pm0.002$. We can
also estimate when the first common horizon forms by calculating the
offset between the tallest peak in the plunge waveform for these data
and that of the 'ISCO' (as determined by the effective potential
method) initial data. In a previous paper~\cite{Campanelli:2005dd} we
found that a common horizon forms in the 'ISCO' case at
$t=19.333M$. From these offsets we estimate that the first common
horizon should form at $111.5M$ for the $M/21$ run, $112.9M$ for the
$M/24$ run, and $113.7M$ for the $M/27$ run. These numbers are within
$0.6M$ agreement with those determined directly from the apparent
horizon finder. An extrapolation of these estimates for the formation
of the common horizon yields $T_{cah} = 121.5M$.

We calculated the radiated energy and angular momentum from $\psi_4$
at $r=15M$, $r=20M$, $r=25M$, and $r=30M$. We then extrapolated these
data (based on a least squares fit versus $\rho = 1/r$) and found that
extrapolated radiated energy and angular momentum were
$(3.18\pm0.2)\%$ and $(24.3\pm2)\%$ respectively. Notably, these
results are in excellent agreement with those calculated from the
horizon mass and angular momentum.  Table~\ref{table:results}
summarizes the physical parameters extracted independently from the
horizon and waveform.

\begin{figure}
\begin{center}
\includegraphics[width=3.0in]{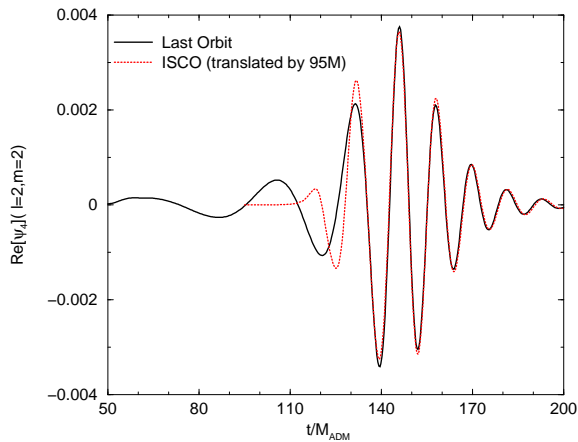}
\caption{The real part of the $(\ell=2,m=2)$ mode of the $\psi_4$ at
$r=20M$ from this `last orbit' configuration and from the ISCO
configuration.
Note the near perfect overlap once the ISCO waveform
has been translated by $\Delta t/M = 95$
}
\label{fig:psi4_compare}
\end{center}
\end{figure}

The plunge part of the waveform shows remarkable similarities with
the waveform from the ISCO. Figure~\ref{fig:psi4_compare} shows the
real part of the $(\ell=2,m=2)$ mode for these two configurations.
Note that, after a translation, there is near perfect overlap of the
late-time waveform.

\begin{figure}
\begin{center}
\includegraphics[width=3.0in]{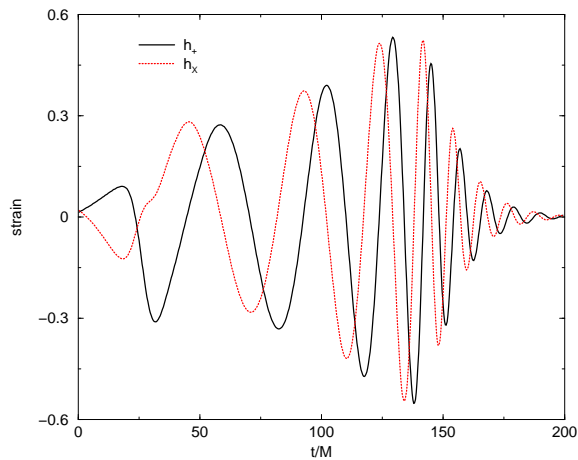}
\caption{The $(\ell=2,m=2)$ component of the strain. Both the $+$ and
$\times$ mode are shown. The early time strain is dominated by spurious
radiation (from the initial data) up to $t=55M$.  Afterwords, the strain
shows a gradual transition from orbital motion to a plunge waveform.
This transition is less distinct that that in $\psi_4$.
}
\label{fig:strain}
\end{center}
\end{figure}

The gravitational strain $h$ is related to $\psi_4$ by $\psi_4 = -
1/2\, \partial_t^2 h(t)$.  In Fig.~\ref{fig:strain} we show the
$(\ell=2,m=2)$ component of both polarizations of the strain.  The
early part of the strain ($t<55M$) is dominated by the spurious
radiation of the initial data. Note that the strain amplitude and
frequency shows a far more gradual transition from an orbital inspiral
type waveform to a plunge type waveform than $\psi_4$ and seems better
suited to match to post-Newtonian waveforms.

\begin{figure}
\begin{center}
\includegraphics[width=3.0in]{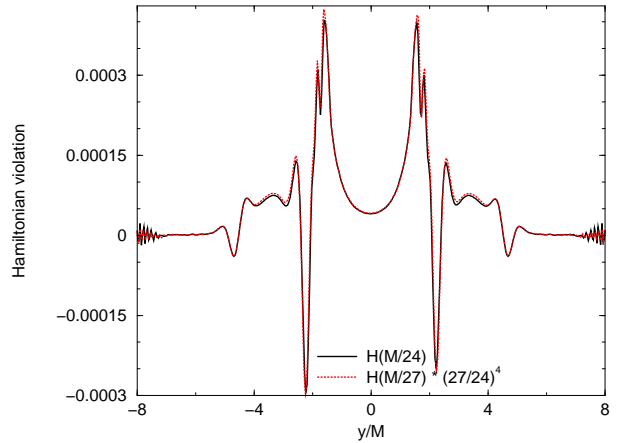}
\caption{The Hamiltonian constraint violation at $t=70M$ along the
y-axis for the $M/24$ and $M/27$ runs (the latter rescaled by
$(27/24)^4$. The punctures crossed the y-axis for the second time at
$t=64M$.  Note the near perfect fourth-order convergence. Points
contaminated by boundary errors have been excluded from the plot.  The
high frequency violations near the numerical coordinate $y=\pm8$ are
due to the extreme fisheye deresolution near the boundary, and
converge with resolution.  }
\label{fig:HC_conv}
\end{center}
\end{figure}

To demonstrate consistency with the General Relativity field
equations, we calculated the Hamiltonian constrain violation. The
constraint converges to fourth-order outside of a small region
surrounding the puncture (the Hamiltonian constraint violation on the
nearest neighboring points to the puncture are roughly independent of
resolution, but this non-converging error does not propagate outside
the individual horizons). Figure~\ref{fig:HC_conv} shows the
Hamiltonian constraint violation along the $y$-axis at $t=70M$ (about
$4M$ after the punctures cross the $y$ axis for the second time). The
constraint is fourth order convergent everywhere but at points
contaminated by boundary errors (these points have been removed from
the plot).

\begin{table}
\caption{Results of the evolution as determined from the waveform and
the remnant horizon.}
\begin{ruledtabular}
\begin{tabular}{lllll}\label{table:results}
Method & $E_{rad}/M_{ADM}$ & $J_{rad}/J_{ADM}$ & $T_{cah}/M$ &
$a/M_{\cal H}$ \\
\hline
Radiation & $(3.18\pm0.2)\%$ & $(24.3\pm2)\%$ & $\approx 121$ &
$0.673\pm0.002$ \\ Horizon & $(3.3\pm0.2)\%$ & $(24.7\pm0.4)\%$ &
$\approx125$ & $0.688\pm0.001$ \\
\end{tabular}
\end{ruledtabular}
\end{table}

\section{Discussion}\label{Sec:Dis}

Using our new technique that allows punctures to move in the numerical
grid, we have succeeded in accurately computing the last orbit of a
black-hole binary, obtained waveforms and extracted relevant physical
information such as energy and angular momentum radiated, apparent
horizon geometry, and orbital parameters. These results are consistent
with each other, as summarized in Table
\ref{table:results}. We also note the interesting fact that the plunge
part of the waveforms (corresponding to the highest amplitude region in
$\psi_4$) is roughly insensitive to the initial separation of
the holes when starting from a quasicircular orbit (see
Fig. \ref{fig:psi4_compare} here and Fig. 30 in Ref. \cite{Baker:2002qf})

When evolving the same initial configuration, Br\"ugmann et.\
al.~\cite{Bruegmann:2003aw} did not find a common horizon (they
evolved to $185M$), and concluded that the binary must have undergone
at least one orbit since the orbital period of the initial
configuration is around $120M$. Recently, Diener et.\
al.~\cite{Diener:2005mg} evolved this system with various choices for
the gauge (all containing a corotating shift) and concluded that a
common horizon forms at about $120M$ - $125M$ (after the binary
completes a full orbit).  They get these estimates by evolving with
the extremely high resolutions of $M/66$ and $M/80$, and extrapolating
to the continuum limit.  They compute the trajectory of the apparent
horizons and find that the orbital period of the last full orbit is
$59M$.  In both~\cite{Bruegmann:2003aw} and~\cite{Diener:2005mg}
waveform extraction was not possible.  In our work we find, like
Diener et.\ al., that the common horizon forms between $120M$ and
$125M$, and that the orbital period of the last orbit is approximately
$62M$. However, we only required a resolution as high as $M/27$, and
could calculate accurate waveforms.

Aside from the duration of the last orbit, we estimate that the
initial separation of the black holes in the final orbit 
is $5.2M$ in conformal
coordinates. It is interesting to compare this separation
with those of several ISCO determinations. For Bowen-York initial data
(as was used in this paper) the ISCO separation~\cite{Cook91,Baumgarte00a} 
is  $2.3M$, for Thin-Sandwich data~\cite{Cook:2004kt,Hannam:2005rp}
it is $3.25M$, and an  estimate of the 3PN~\cite{Baker:2002qf} ISCO
puts the separation at $4.24M$.  Obviously,
radiation reaction, which was not taken into account in those
computations, leads to radial motion that plays an important role in the
dynamics of the last orbit of black-hole binaries.

\acknowledgments
We thank Erik Schnetter for providing the thorns to implement
Pi-symmetry boundary conditions.  We gratefully acknowledge the
support of the NASA Center for Gravitational Wave Astronomy at
University of Texas at Brownsville (NAG5-13396) and the NSF for
financial support from grants PHY-0140326 and
PHY-0354867. Computational resources were performed by the 64-node
Funes cluster at UTB.
	
\bibliographystyle{apsrev}
\bibliography{../../Lazarus/bibtex/references}
\thebibliography{rp}

\end{document}